\documentclass[12pt]{JHEP}
\usepackage{amsmath,amsfonts,amssymb,amsthm,amstext,amscd}

\newcommand{\be}{\begin{equation}}
\newcommand{\ee}{\end{equation}}
\newcommand{\beqa}{\begin{eqnarray}}
\newcommand{\eeqa}{\end{eqnarray}}

\def\med{\frac{1}{2}}

\def\Gmeas{\sqrt{-G}}

\def\d{\partial}
\def\pp{{\cal P}}

\def\o{\omega}

\def\b{\beta}

\def\dw{\mathfrak{w}}
\def\dq{\mathfrak{q}}

\relax

\title{Hydrodynamics from  the Dp-brane}

\author{ Javier Mas and Javier Tarr\'\i o 
\\

 Departamento de F\'\i sica de Part\'\i culas,
Universidad de Santiago de Compostela \\ E-15782 Santiago
de Compostela, Spain
}

\medskip

\abstract{We complete the computation of viscous transport coefficients in the near horizon geometries that arise from a stack of black Dp-branes for $p=2,...,6$ in the decoupling limit. The main new result is the obtention of the  bulk viscosity which, for all $p$, is found to be related to the speed of sound  by the   simple relation $\zeta/\eta = -2(v_s^2-\frac{1}{p})$. For completeness 
 the shear viscosity is rederived from gravitational perturbations in  the shear and scalar channels.
We comment on technical issues like the counterterms needed, or the possible dependence on the conformal frame.}
\keywords{AdS/CFT, Hydrodynamics}

\begin{document}

\section{Introduction and statement of results}
\label{secintro}
The AdS/CFT correspondence is believed to apply  to stacks of Dp-branes for arbitrary $p$ \cite{Itzhaki:1998dd}. The non-conformality of the Dp-brane backgrounds, being welcome for the physics, introduces a host of technical difficulties. On the side of the boundary theory, the identification as a ``bona fide"  QFT works only within some energy windows. Still a lot of physics has been extracted from such an effective description. For example, for $p>3$, wrapping some internal   world-volume directions of the brane along a small enough compact manifold has become an industry for modelling supersymmetric versions of QCD, starting with D4-branes in \cite{Witten:1998zw},  D5-branes in \cite{Maldacena:2000yy} and D6-branes in \cite{Edelstein:2001pu}.

The UV completion of these theories gives in most cases an elusive object. On the gravity side, this translates into the absence, so far, of a full fledged holographic renormalization program, as complete as the one developed for asymptotically AdS metrics (see \cite{Skenderis:2002wp} and references therein).  In the case of Dp-branes, the metric in the decoupling limit is only conformal to AdS$_{p+2}$. Still a minimally modified  set of counterterms was proposed in \cite{Cai:1999xg} to renormalize the on-shell boundary  action.  This, by itself, sets the thermodynamics under control  and allows for the computation of the energy-momentum tensor, which matches  the one obtained from the asymptotically flat completion  \cite{Myers:1999ps}.  In this paper we will see  that such counterterms are also enough to obtain the shear viscosity from the two point function of the energy-momentum tensor. A full construction of the renormalized action is clearly beyond the scope of this  note, and should presumably proceed along the lines investigated in  \cite{Papadimitriou:2004ap}.

In this paper we will study the transport coefficients of the dual plasma in the universal hydrodynamic regime.
This implies that all time-lenght scales have to be very large as compared to the microscopic correlation lenghts, which are set by the  inverse temperature $T^{-1}$.
Having control over the thermodynamics gives already information about one transport coefficient, namely the speed of sound  $v_s^2 = \d P /\d \epsilon$. The following expression
\be
v_s^2 = \frac{5-p}{9-p}\label{spos}
\ee
albeit  evident from the form of the renormalized energy-momentum tensor  \cite{Cai:1999xg,Myers:1999ps}, was to our knowledge first written in \cite{Caceres:2006ta}. It  signals the onset of a tachyonic instability that is dual to the fact that for $p>5$ the specific heat becomes negative \cite{Buchel:2005nt}. 

 Among other results in this paper, we will  recover \eqref{spos} from the pole structure of retarded correlators
 of the energy-momentum tensor. The implementation of this program in the context of AdS/CFT correspondence
 was initiated in \cite{Policastro:2002se,Policastro:2002tn} and we will make use of the clean formulation advocated in \cite{Kovtun:2005ev} that neatly explains how to obtain the relevant dispersion relations from gauge invariant fluctuations of the supergravity fields. The key observation is the fact that the relevant boundary conditions for the fluctuations
are the same as the so called quasi-normal modes in the context of black hole perturbation analysis.
Quasinormal modes for p-branes  have been studied in the past, albeit in different context. In \cite{Iizuka:2003ad, Maeda:2005cr}  the emphasis was on the thermalization properties of the dual plasma.
In \cite{Hoyos:2006gb,Mateos:2007vn} the aim was to investigate the decay of probe-branes in a thermal AdS background.

In contrast, hydrodynamics is related to the long wavelength/frequency limit of  perturbations, hence to the  lowest such quasinormal modes. 
Symmetry analysis allows to catalog the fluctuations in three decoupled channels. In two of them, so called shear and sound channels, the general formalism predicts the appearance of poles of the following form
\beqa
\hbox{shear channel} & \to & \o = -\frac{i\eta}{\epsilon+P}q^2,\label{dispshear}\\
\rule{0mm}{7mm}
\hbox{sound channel} & \to & \o = v_s q-i\frac{\eta}{\epsilon+P}\left(\! \frac{p-1}{p}+\frac{\zeta}{2\eta}\!\right)q^2 + \cdots \label{dispsound}
\eeqa
Microscopically, such dispersion relations turn into poles of the retarded two point functions of certain components of the energy-momentum tensor. The relevant two point functions were precisely identified in \cite{Kovtun:2005ev} with 
fluctuations of the background metric that transform respectively as a vector and a scalar under the little group $SO(p-1)$ (resp. shear and sound channels).
As we will show below, the dispersion relations  allow to recover both the speed of sound given in \eqref{spos} as well as the shear and bulk viscosities. The results are best expressed in terms of the
following quotients
\be
\frac{\eta}{s} = \frac{1}{4\pi}, \hspace{1cm} \frac{\zeta}{\eta} =  \frac{2(3-p)^2}{p(9-p)}~.
\label{results}
\ee
The famous equation on the left hand side was first obtained in \cite{Kovtun:2003wp} both in the context of the membrane paradigm and in the AdS/CFT formalism by relating  the shear viscosity with the diffusion of the R-current. Just for completeness, we add here a genuinely gravitational computation.

 Also following an observation of  \cite{Benincasa:2006ei} we notice that 
from equations \eqref{dispshear} and \eqref{dispsound}  the relation
\be
\frac{\zeta}{\eta}=-2\left(v_s^2-\frac{1}{p}\right) \label{vsbulk}
\ee
 holds exactly true for all values of $p$. We will have more to say about this equation in the conclusions.
 
  The paper is organized as follows. In section 2 we shall establish the reduced model in the $p+2$ dimensional bulk, and argue that it only contains a scalar field in addition to the metric. In the next section we shall examine the fluctuations and obtain the transport coefficients announced in this introduction.
We add a short section which starts by raising the question about the correct choice of  conformal frame.  Unfortunately, the final results for the transport coefficients exhibit no dependence on the frame, and thus, shed no further  light into the question. The paper closes with some concluding remarks and comparison with related results in the literature.

\section{Consistent reduction and thermodynamics}
In the Einstein frame, the relevant supergravity field profiles that correspond to the decoupling limit of a stack of Dp-branes read as follows
\beqa
ds_{10}^2 &=& G^{(10)}_{MN} dx^M dx^N \nonumber\\
&=& 
H^{-\frac{7-p}{8}}(r)(-f(r) dt^2 + dx_1^2 + \cdots + dx_p^2) +  
H^{\frac{p+1}{8}}(r)\left( \frac{dr^2}{f(r)} + r^2 d\Omega_{8-p}^2\right), \label{pmetric}\\
e^{\phi(r)} &=& H(r)^{\frac{3-p}{4}}, \label{dilaton}
\\
F_{(8-p)}&=& \frac{7-p}{L} {\omega}_{S_{8-p}},
\eeqa
where  $H(r) = \left(L/r\right)^{7-p} $,
$f(r) = 1- \left(r_0 /r \right)^{7-p},$ and $d\Omega_{8-p}^2$ stands for the 
metric of a $8-p$ sphere of unit radius.
This solution is obtained from a type II supergravity lagrangian where, keeping only the relevant degrees of freedoms, we have
\be
S_{II}=\frac{1}{16\pi G_{10}}\int d^{10}x\Gmeas\left[R(G)-\med\d_M\phi\d^M\phi-\frac{1}{2 n!}e^{ a\phi}   F_{(8-p)}^2\right],\label{Maction}
\ee
with $a= (3-p)/2$ ({\em i.e.} we are considering magnetically charged branes).
Consider the following ansatz for a dimensional reduction
\beqa
ds_{10}^2 &=&  e^{-\frac{2(8-p)}{p} B(r)} g_{\mu\nu}(x) dx^\mu dx^\nu +  e^{2B(r)} L^2 d\Omega^2_{8-p}\\
&=& 
  e^{-\frac{2(8-p)}{p} B(r)} \left( -c_T^2(r) dt^2 + c_X^2(r) \sum_{i=1}^p dx^2_i
+ c_R^2(r)dr^2 \right) + e^{2B(r)}L^2 d\Omega^2_{8-p},
\label{bulksound}
\eeqa
where $g_{\mu\nu},\, \mu,\nu=0,1,\dots,p$ stands for  the metric in the $p+2$ dimensional Einstein frame.
Plugging this ansatz into the equations of motion derived from  (\ref{Maction}) one
obtains a coupled system of differential equations for the metric components $c_T(r), c_X(r)$
and $c_R(r)$, as well as for the dilaton $\phi(r)$ and the breathing mode $B(r)$.
One can then check easily that the following identification yields a consistent reduction
\be
B(r) = -\frac{3-p}{4(7-p)} \phi(r)\, . \label{reduc}
\ee
 By this we mean that the equations of motion for $B(r)$ and $\phi(r)$, and perturbations thereof, exactly merge. Therefore, from here on we shall work within this truncation. The effective system  in $p+2$ spacetime
is governed by a  system of equations that can be derived from the following action \cite{Boonstra:1998mp}
\be
I_{bulk} = \frac{1}{16 \pi G_{p+2}} \int d^{p+2}x \sqrt{-g} \left(R(g) - \frac{\beta}{2} \d_\mu\phi \d^\mu\phi - \pp(\phi)\right), \label{redac}
\ee
with
$$
\frac{1}{G_{p+2}} =  \frac{ 2\pi^{\frac{9-p}{2}}L^{8-p}}{\Gamma\left(\frac{9-p}{2}\right)G_{10}},
$$
where $\pp(\phi)$ is the effective potential for the dilaton, and we have not bothered to normalize the field $\phi$ canonically
\be
\pp(\phi)=-\frac{(7-p)(p-9)}{2L^2}e^{\frac{4(3-p)}{p(7-p)}\phi(r)}, \hspace{1cm} 
\beta = \frac{8(9-p)}{p(7-p)^2}.
\label{potential}
\ee
The effective equations of motion 
\beqa
R_{\mu\nu}(g) &=&\frac{\beta}{2} \d_\mu\phi\d_\nu\phi+\frac{1}{p}g_{\mu\nu}\pp(\phi),
\label{emg}
\\
\Box\phi&=&\frac{\pp'(\phi)}{\beta} \label{emp} ,
\eeqa
are  satisfied by the background profiles (setting $l_s=1$)
\be
c^2_T(r) =\left(\frac{r}{L}\right)^{\frac{9-p}{p}} f(r), \hspace{1cm} 
c^2_X (r)=\left(\frac{r}{L}\right)^{\frac{9-p}{p}}, \hspace{1cm} 
c^2_R(r) = \frac{1}{f(r)}\left(\frac{r}{L}\right)^{\frac{p^2-8p+9}{p}},
\label{backg}
\ee
as well as 
\be
\phi(r) =- \frac{(3-p)(7-p)}{4}\log \left(\frac{r}{L}\right) \, , \label{backphi}
\ee
from where the Hawking temperature and entropy density come  straight  
\be
T = 	\frac{7-p}{4\pi r_0} \left( \frac{r_0}{L}\right)^{(7-p)/2}, \hspace{1cm} 
s = \frac{1}{4 G_{p+2}} \left( \frac{r_0}{L}\right)^{(9-p)/2}\label{temps}\, .
\ee
If desired, using the AdS/CFT dictionary, it is straightforward to translate the entropy density into   field theoretical quantities, involving the rank $N$, the
temperature $T$, and  the gauge coupling $\lambda= N g_{YM}^2 = (2\pi)^{p-2} g_s l_s^{p-3}$ 
\be
s \sim N^2 \lambda^{-\frac{3-p}{5-p}}T^{\frac{9-p}{5-p}}.
\ee
 

\section{Fluctuations in the hydrodynamic regime}
Let us consider fluctuations of the bulk fields
$
g_{\mu\nu} \rightarrow g_{\mu\nu}+\delta g_{\mu\nu} ,~
\phi \rightarrow \phi+\delta\phi, 
$
and focus on a single Fourier component that propagates along the coordinate $z= x^p$
\beqa
\delta g_{\mu\nu}(t,z,r) &=& e^{-i(\o t-qz)}h_{\mu\nu}(r), \\
\delta\phi(t,z,r)  &=& e^{-i(\o t-qz)}\varphi(r).
\eeqa
 Standard analysis proceeds by 
grouping the fluctuations into three  irreducible channels according to their helicity $s$ under the
little group $SO(p-1)$ \cite{Policastro:2002se}
\beqa
s=0&\to&\textrm{sound channel :}~h_{tt},\,h_{tz},\,h_{zz},\,h_{rr},\,h_{tr},\,h_{zr},\,h, \varphi
\\
s=1&\to&\textrm{shear channel :}~h_{ta},\,h_{za},\,h_{ra}
\\
s=2&\to&  \textrm{scalar channel :}~h_{ab}-\delta_{ab}\frac{h}{p-1}
\eeqa
with $a,b=1,\dots, {p-1}$ and $h = \sum_{a} h_{aa}$. Let us  parametrize fluctuations as usual with $H_{\mu\nu}(r)$ such as
\beqa
h_{tt}(r)&=& c_T^2H_{tt}(r),\\
h_{\mu j }(r)   &=& c_X^2H_{\mu j }(r) ,
\eeqa
with $ h_{j \mu }(r) = h_{\mu j}(r)$ and  $ j = 1,2,\dots,p$.
We have fixed coordinates such that $\delta g_{\mu r} = 0$.
This leaves  still a residual gauge freedom under the infinitesimal diffeomorphisms $x^\mu\to x^\mu+\xi^\mu$, $\delta g_{\mu\nu}\to \delta g_{\mu\nu}-\nabla_\mu\xi_\nu-\nabla_\nu\xi_\mu$ and $\delta \phi\to \delta\phi-\partial^\mu\phi\xi_\mu$ with $\xi_\mu = \xi_\mu(r)e^{-i\o t+iqz}$ and covariant derivatives taken with respect to the background metric. Rather than fixing completely the gauge, it is   more convenient to switch over to a set of gauge invariant fluctuations  \cite{Kovtun:2005ev}
\beqa
&& \left\{
\begin{array}{rcl}
Z_0&=&\displaystyle q^2 \frac{c_T^2}{c_X^2}H_{tt}+2q\o H_{tz}+\o^2H_{zz}+\!\left(\!q^2  \frac{\ln'(c_T)}{\ln'(c_X)} \frac{c_T^2}{c_X^2}-\o^2\!\right)\!H,\label{gauge-sound}\\ 
Z_\varphi &=& \displaystyle\varphi-\frac{\phi'}{\ln'(c_X{^{2(p-1)})}}H_{aa},
\end{array}
\right. \label{invsound}
\\
\rule{0mm}{7mm}
&&~~~ Z_1=qH_{ta}+\o H_{za},  \label{infsh}\\
&& ~~~ÊZ_2 =\rule{0mm}{6mm}H_{ab}\, , \label{invscal}
\eeqa
where $H=\frac{1}{p-1}\sum_a H_{aa}$.
The ODEs obeyed by the fluctuations $H_{\mu\nu}$ can be found in appendix \ref{Ap:pert}. Here we just present the equations for the gauge invariant fluctuations.
The following dimensionless ratios are natural in order to examine the hydrodynamic regime
\be
\dw=\frac{\o}{2\pi T}, \hspace{1cm} \dq=\frac{q}{2\pi T}\, .
\ee
In each one of the three channels we shall obtain decoupled second order differential equations that will be solved in the nontrivial lowest order limit when $\dw \ll 1$ and $\dq \ll 1$ with $\dw/\dq = \lambda(\dq) $, where $\lambda(\dq)$ is a function of $\dq$ analytic at $\dq \to 0$. The analysis of the characteristic exponents near $r_0$ allows us to parametrize our gauge invariant functions as follows
\be
Z_{x}(r)=f(r)^{-i\frac{\dw}{2}} Y_x(r), \label{ingoing}
\ee
with $Y_x(r)$ analytic at $r=r_0$. In this way we are selecting ingoing boundary conditions at the horizon. Then we only have to solve  perturbatively for $Y_x(r)$ in the hydrodynamic limit.
The dispersion relation is obtained from imposing Dirichlet boundary conditions 
\cite{Kovtun:2005ev}
\be
\left. Z_x(r)\right\vert_{r=\infty} = 0 . \label{disprel}
\ee

\subsection{Shear Channel}
Taking a suitable  combination of the three equations  in section \ref{Ap:shear} one  obtains the ODE satisfied by the gauge independent vector fluctuation $Z_1$
\be
Z_1''+\!\left[\left(\frac{q^2}{\o^2}\frac{c_T^2}{c_X^2}-1\right)^{-1}\!\ln'\!\left(\frac{c_X^2}{c_T^2}\right)+\ln'\!\left(\!\frac{c_X^{p+2}}{c_Tc_R}\!\right)\!\right]\!Z_1'+c_R^2\!\left(\!\frac{\o^2}{c_T^2}-\frac{q^2}{c_X^2}\!\right)\!Z_1=0.\label{sheq}
\ee
Plugging the ingoing ansatz (\ref{ingoing}) we obtain an equation for $Y_1(r)$ which can be solved perturbatively giving in all cases
\be
Z_1(r) = C_V f(r)^{-i\frac{\dw}{2}}\left(1+i\frac{\dq^2}{2\dw}f(r)+{\cal{O}}(\dw,\dq^2)\right),
\ee 
where $C_V$ is an unimportant normalization factor.
From here and (\ref{disprel})  the familiar dispersion relation follows
\be
\dw =-i \frac{\dq^2}{2}.
\ee
Restauring $\omega$ and $q$, and comparing  with  \eqref{dispshear},  gives  the standard result  
for $\eta/s$ shown in the left hand equation of \eqref{results}.


\subsection{Sound Channel}

In this channel we end up with two scalar fluctuations $Z_\varphi$ and $Z_{0}$.
They fulfill the following equations
\beqa
&& Z_\varphi''+\ln'\!\left(\!\frac{c_Tc_X^p}{c_R}\!\right)\!Z_\varphi'+c_R^2\left[\frac{\o^2}{c_T^2}-\frac{q^2}{c_X^2}-\frac{2(3-p)}{p}\left(\frac{3-p}{9-p}+\frac{\phi'}{\ln'(c_X)}\frac{2}{(7-p)p}\right)\pp\right]Z_\varphi=0	\label{eqzphi}
\\
\rule{0mm}{6mm}
&& Z_0''+{\cal{F}}(r)Z_0'+{\cal{G}}(r)Z_0+{\cal{H}}(r)Z_\varphi=0\, ,\label{gaugesound}
\eeqa
modulo the background equations of motion.
The coefficients in \eqref{gaugesound} are given by
\beqa
{\cal{F}}(r)&=&\ln'\!\left(\!\frac{c_Tc_X^p}{c_R}\!\right)\!-4\ln'\left(\frac{c_T}{c_X}\right) +  \xi(r),\\
{\cal{G}}(r)&=&c_R^2\!\left(\!\frac{\o^2}{c_T^2}-\frac{q^2}{c_X^2}\!\right)\!+4\left[\ln'\left(\frac{c_T}{c_X}\right) \right]^2 -  \ln'\left(\frac{c_T}{c_X}\right) \,\xi(r),\\
{\cal{H}}(r)&=&8\frac{q^2}{\o^2}\frac{(3-p)}{p(p-7)} \frac{c_T}{c_X^2} \left(c_T'-c_T\ln'(c_X)\right) \,\xi(r),\label{hache}
\eeqa
with
\be
\xi(r)=\frac{q^2 (c_T^2)' \frac{\ln''(c_X)}{(\ln'(c_X))^2} \left( 1 - \frac{\ln''(c_T) \ln'(c_X)}{\ln''(c_X) \ln'(c_T)} \right) + 2 \o^2 p (c_X^2)' \left( 1 - \frac{\ln'(c_T)}{\ln'(c_X)} \right)}{q^2 c_T^2 \left(\frac{\ln'(c_T)}{\ln'(c_X)} + p - 1 \right) - \o^2 p c_X^2}.
\ee
Plugging as before the ingoing ansatz (\ref{ingoing}) and solving perturbatively  one finds that the only non-singular  solution to \eqref{eqzphi} is a constant, which we set to zero by the boundary conditions at infinity. Inserting now
$Z_\varphi = 0 $ into \eqref{gaugesound} and solving perturbatively for $Z_0$ gives
\be
Z_0(r) = C_S f(r)^{-i\frac{\dw}{2}}\left(1-\frac{(1+i4\dw)\dq^2f(r)}{(7-p)\dq^2-(9-p)\dw^2}+{\cal{O}}(\dw^2,\dq^2,\dw \dq)\right).\\
\ee
with $C_S$ a normalization constant.
Imposing the Dirichlet boundary condition (\ref{disprel}) gives an expression  for $\dw(\dq)$ that
can be expanded as follows
\be
\dw = \sqrt{\frac{5-p}{9-p}}\,\dq-i\frac{2}{9-p}\dq^2+\cdots\label{disprel2}
\ee
Comparing this expression with the dispersion relation \eqref{dispsound}, and using \eqref{results} we identify finally
\be
v_s^2=\frac{5-p}{9-p}\, , \hspace{1.5cm}\frac{\zeta}{\eta}= \frac{2(3-p)^2}{p(9-p)}\, ,  \label{sspeed}
\ee
as claimed in the introduction.

\subsection{Scalar Channel}

In this subsection, and just for the sake of completeness,  we reobtain  the shear viscosity through the Kubo formula. It is little more than an academic exercise, given the general theorem \cite{Buchel:2004qq}. However the fact that the metric is not asymptotically AdS makes it worth  to explore this in detail.
As usual, the equation satisfied by $Z_2=H_{ab}$  is   that of a minimally coupled scalar 
\be
Z_2''+\ln'\!\left(\!\frac{c_Tc_X^p}{c_R}\!\right)\!Z_2'+c_R^2\!\left(\!\frac{\o^2}{c_T^2}-\frac{q^2}{c_X^2}\!\right)\!Z_2=0. \label{eqzdos}
\ee
 In the hydrodynamic limit the ingoing solution to \eqref{eqzdos} exhibits no poles and can be expanded as follows
    \be
  Z_2(\omega,r) =   f(r)^{-i\frac{\dw}{2}}(1 + {\cal O}(\dw^2, \dq^2)).
  \ee
In this case the standard roundabout invokes the Kubo formula
\be
\eta = - \lim_{\omega\to 0} \frac{1}{\omega} {\rm Im} \, G_R(\omega),
\ee
with $G_R$ the retarded correlator of the relevant components of the energy-momentum tensor
\be
G_R(\omega) = -i \int  dtd^{p}x  e^{i\omega t} \theta(t) \langle [ T_{xy}(t,\vec x),T_{xy}(0,\vec 0) ] \rangle.
\ee
The evaluation of the retarded correlator calls for the expansion of the renormalized boundary action up to second order in the fluctuation $H_{\mu\nu}$. Whereas such an object has been rigurously constructed for actions possesing asymptotically locally AdS$_d$ backgrounds (see \cite{Skenderis:2002wp} and references therein), for the case of Dp-branes
only partial results are known. In \cite{Cai:1999xg} appropriate counterterms where found
on a case by case basis that properly renormalized the action, giving a renormalized energy-momentum tensor. One can easily see that these counterterms, with the correct coefficients, are exactly reproduced by  the general expresion given in  \cite{Batrachenko:2004fd} which we follow here. Let us express the  regularized action as
\be
I = \sum_A  I_A = I_{bulk} + I_{GH} + I_{ct},
\ee
where $I_{bulk}$ is as in \eqref{redac}, and 
\be
I_{GH} = \frac{1}{16\pi G_{p+2}}\int_{r=r_\infty} d^{p+1}x \sqrt{h}\,  2K, \hspace{1cm} 
I_{ct} = \frac{1}{16\pi G_{p+2}}\int _{r=r_\infty}d^{p+1} x \sqrt{h}\left( \,2{\cal  W}(\phi) +\cdots\right).
\label{regac}
\ee
${\cal W}(\phi)$ is the superpotential, related to the potential ${\cal P}(\phi)$ by the non-linear equation
\be
{\cal P}(\phi) = \frac{2}{\beta} (\partial_\phi{\cal W}(\phi))^2-\frac{p+1}{p} {\cal W}(\phi)^2,
\ee
whose solution, for ${\cal P}(\phi)$ as in \eqref{potential},  is given by 
\be
 {\cal W}(\phi) =
 \frac{(9-p)}{2L}e^{\frac{2(3-p)}{p(7-p)}\phi(r)}.
\ee
The dots in \eqref{regac} denote higher curvature invariants on the induced hypersurface. After expanding $I$ to second order in the (purely time dependent) perturbation
\be
h^a{_b}(t,r) =\int \frac{d\omega}{2\pi} \,e^{i\omega t} f(\omega)Z_2(\omega,r),
\ee
we can cast all contributions in the form of boundary terms $\delta I = \sum_A \delta I_A$ with
\be
\delta I _{A} = \left. \int d^p x  \frac{d\omega }{2\pi}   f(\omega) f(-\omega) {\cal F}_A(\omega, r)\right\vert_{r_0}^\infty
\ee
and find thereafter
\beqa
{\cal F}_{bulk} &=& \frac{1}{16\pi G_{p+2}}\left(\frac{r_0^{7-p}}{L^{8-p}}\right)
\left(-\frac{9-p}{2p}\left( \frac{r}{r_0} \right)^{7-p} +\frac{9-p}{2p}+ i \frac{3(7-p)}{4} \dw + ...\right),
\nonumber\\
{\cal F}_{GH} &=&\frac{1}{16\pi G_{p+2}} \left(\frac{r_0^{7-p}}{L^{8-p}}\right)
\left( \frac{(9-p)(p+1)}{2p} \left( \frac{r}{r_0} \right)^{7-p} - \frac{9+p}{2p} - i (7-p)\dw +  ...\right), \nonumber\\
{\cal F}_{ct} &=& \frac{1}{16\pi G_{p+2}}\left(\frac{r_0^{7-p}}{L^{8-p}}\right)
\left(  -\frac{9-p}{2} \left( \frac{r}{r_0} \right)^{7-p} + \frac{9-p}{4} + ...\right),
 \nonumber
\eeqa
where the dots stand for terms of $ {\cal O}(r_0/r, \dw^2) $.
Adding up and using the Minkowskian prescription of \cite{Son:2002sd} we obtain the retarded correlator
\be
G_R(\omega)= \left. 2 {\cal F}(r)\right\vert_{r=\infty} = \frac{1}{16\pi G_{p+2}} 
\left(\frac{r_0^{7-p}}{L^{8-p}}\right) \left(
\rule{0mm}{4mm}\frac{5-p}{2} -  i  \frac{7-p}{2}\dw\right). \label{eqgr}
\ee
We see that the counterterm contributes to the real part of the renormalized correlator\footnote{ Such contribution is essential in order to fulfill Ward identities, and is tipically missed in non-covariant treatments of the counterterm action.  It does not contribute
to the coefficient  of $\omega$ but, to our knowledge, this is not a general statement when higher curvature counterterms are included.}. From \eqref{eqgr}  the shear viscosity can be extracted as usual by means of Kubo formula
\be
\eta = \frac{1}{16\pi^2 G_{p+2} } \frac{7-p}{4}\frac{r_0^{7-p}}{TL^{8-p}}
\ee
 and, using \eqref{temps}, again the well known result  $\eta/s = 1/4\pi$ is recovered.

\section{Frame (in)dependence}

In the  AdS/CFT correspondence, one delicate issue concerns  the correct identification of the bulk field perturbation that couples correctly to the desired boundary operator.
We are interested in perturbing bulk fields that couple exactly to the boundary energy-momentum tensor. In asymptotically AdS spaces, the energy-momentum tensor couples naturally to the bulk metric. However, outside the well tested arena of such backgrounds, we are on less firm grounds. In the case of Dp-branes, the metric is asymptotically conformally AdS$_{p+2}$ in the Einstein frame\footnote{except for $p=5$ where it asymptotes to flat $\mathbb{M}^{1,6}$}. The conformal factor that asymptotically deviates its profile from AdS  is 
given by an appropriate power of the function  $e^{\phi(r)}$ where $\phi(r)$ is the dilaton. Hence we may consider   a family of conformally related metrics,  parametrized 
by $\alpha\in  \mathbb{R} $ as follows
\be
 g_{\mu\nu}  = e^{2\alpha\phi }g_{\mu\nu}^{(\alpha)}.  
\ee
Clearly $g^{(0)}_{\mu\nu}$ is the Einstein frame metric whose background value is given in \eqref{backg}. Another important case yields the so called ``dual frame" \cite{Boonstra:1998mp}, and is obtained by tuning $\alpha$ to the following value
$$
\alpha_D= -\frac{2(3-p)}{p(7-p)},
$$ 
which sets exactly  $g_{\mu\nu}^{(\alpha_D)} $ to the following asymptotically AdS$_{p+2}$ black hole metric
\be
ds^2_{(\alpha_D)} = \left(\frac{r}{L}\right)^{5-p}(-f(r) dt^2 + d\vec x_p d\vec x_p) + \left(\frac{L}{r}\right)^2 \frac{dr^2}{f(r)}.
\ee
 In \cite{Boonstra:1998mp}, this conformal frame was argued to yield the natural ``holographic"  bulk metric, where the AdS/CFT duality should  work   most transparently.
Notice that, in principle,  perturbation of the metric in different frames would couple to different combinations of the energy-momentum tensor and the ``glueball operator" in the boundary field theory.
Until the question among the dual frame and the Einstein frame is settled, 
the   natural way to proceed is   to see if indeed the results depend upon the choice of such frame.
We can repeat the analysis of the paper in terms of the pair $(g_{\mu\nu}^{(\alpha)} ,\phi)$
for an arbitrary $\alpha$. In particular this amounts to replacing $g_{\mu\nu} =
 e^{2\alpha\phi } g_{\mu\nu}^{(\alpha)}$ in equations \eqref{emg} and \eqref{emp} . Introducing now perturbations as follows  
 \beqa
g^{(\alpha)}_{\mu\nu}&\rightarrow&g^{(\alpha)}_{\mu\nu}+\delta g^{(\alpha)}_{\mu\nu},\\
\phi&\rightarrow&\phi+\delta\phi,
\eeqa
 all the intermediate equations acquire a dependence on $\alpha$. For example, instead of 
\eqref{eqzphi} one gets 
\beqa
Z_\varphi'' &+& \left(\ln'\!\left(\!e^{p\alpha\phi}\frac{c_Tc_X^p}{c_R}\!\right)\right)\!Z_\varphi'\nonumber\\
&+&c_R^2\left[\frac{\o^2}{c_T^2}-\frac{q^2}{c_X^2}-\frac{2(3-p)}{p}\left(\frac{3-p}{9-p}+\frac{\phi'}{\ln'(c_X)}\left( \frac{2}{(7-p)p} +\alpha\frac{3-p}{9-p}\right)\right)\pp \right]Z_\varphi=0
\eeqa
and so on. The analysis can be carried along the same lines as in before, and  in the final
result all the $\alpha$ dependence cancels out exactly. Stated precisely, the expressions
given in \eqref{spos} and \eqref{results} are frame independent.

\section{Conclusions}
In this note, we have completed the table of fluid transport coefficients of the non-abelian 
quantum plasmas that are dual to the gravitational background of a stack of non-extremal  Dp-branes in the decoupling limit for $p=2,...,6$.  We have recovered known values for the speed of sound \eqref{spos} and the quotient of the shear viscosity over the entropy \eqref{results} from poles of energy-momentum tensor correlators, as well as from the Kubo formula. The main new result is the expression for the bulk viscosity given in \eqref{results}, which leads to the compact relation \eqref{vsbulk}. Besides we have clarified some aspects related to the holographic renormalization and the frame dependence of the metric.

Let us comment and compare with partial results obtained in the literature in similar contexts. 
In \cite{Parnachev:2005hh}, Parnachev and Starinets also investigate the hydrodynamic properties of thermal ``little string theory" (LST), which is dual to a stack of black NS5-branes. Their results can be seen to agree with ours for $p=5$, reflecting the fact that viscosity (as it happens with  the entropy) is  an S-duality invariant. 

 In  reference  \cite{Benincasa:2006ei}, Benincasa and Buchel consider the background of a stack of D4-branes with one compactified dimension. This seemingly complicated geometry led them to introduce  up to three independent scalar modes. 
Finally the dispersion relation they obtain matches precisely  our eq. \eqref{disprel2} with $p=4$
\footnote{This points to a manifestation of the D4 structure that underlies the construction. However this fact is less than trivial. One can look at the equations for the fluctuations in
\cite{Benincasa:2006ei}, and find that they agree with our equations (both for $H_{\mu\nu}$ in the appendix A, and for  $Z_x$ in the main text) by setting instead $p=3$. Also the gauge invariant combinations are given by \eqref{invsound} with $p=3$. This happens because these expressiones are only sensitive to the value of $p$ that appears in the metric ansatz.  It is only upon inserting the precise values of the field profiles, that the authors of \cite{Benincasa:2006ei}  use the ones in  \eqref{pmetric} and \eqref{dilaton} with $p=4$. Somehow the $p$ in  the final dispersion relation
\eqref{disprel2} refers to its value in the background profile and loses track of the form of the metric ansatz.}.
 The  disagreement comes from extracting  the bulk viscosity, where the authors
of \cite{Benincasa:2006ei}, having in mind a three dimensional fluid, use a parametrization of the dispersion relation which is precisely \eqref{dispsound}  with $p=3$. The obtained value of the bulk viscosity satisfies a relation with the speed of sound which is \eqref{vsbulk} with $p=3$ instead of $p=4$.
Intrigued by this mismatch  we discovered an identity that  extends our equation
\eqref{vsbulk} and encompasses also (1.4) of \cite{Benincasa:2006ei}. Namely, one can replace $p\to d$   in \eqref{dispsound}  and solve again for $v_s$ and $\zeta/\eta$ by comparison with the dispersion relation  \eqref{disprel2} while keeping $p$ and $d$ independent
\be
v_s^2 =\frac{5-p}{9-p}\, , \hspace{1cm} \frac{\zeta}{\eta} = \frac{8d-2(9-p)(d-1)}{d(9-p)}\, .
\label{extres}
\ee
With this,  one can verify that the following identity holds for any $p$ and $d$
\be
\frac{\zeta}{\eta}  = -2 \left( v_s^2 - \frac{1}{d}\right). \label{genvsb}
\ee
Whether this can be ascribed a meaning or happens to be an arithmetic coincidence we don't know yet. Anyway, as announced, the results of \cite{Benincasa:2006ei} are recovered exactly by setting  $d=3$ and $p=4$ in \eqref{extres} and \eqref{genvsb}. Obviously this extension is motivated by the possibility of defining lower dimensional fluids from higher dimensional UV-field theories (e.g. via Kaluza-Klein compactification), but notice  that  it holds as well for $d> p$.
Anyway, the persistence of an analytic pattern like \eqref{vsbulk}, or its generalization \eqref{genvsb}, looks extremely appealing, as it also shows up in the cascading theory \cite{Buchel:2005cv}.  Having encountered some weak deviation in \cite{Benincasa:2005iv}, it does not seem to acquire  yet as universal a character as the famous quotient $\eta/s = 1/4\pi$, but it certainly deserves further attention.

 \section*{Acknowledgments}
 
We would like to thank Alfonso V. Ramallo,  Kostas Skenderis and Andrei Starinets for discussions and comments.
The present work has been supported by MCyT, FEDER  under grant FPA2005-00188, by
 EC Commission under grants HPRN-CT-2002-00325 and MRTN-CT-2004-005104,  and by Xunta de Galicia (Direcci\'on Xeral de Ordenaci\'on e Calidade do Sistema Universitario de Galicia, da Conseller\'\i a de Educaci\'on e Ordenaci\'on Universitaria).


\appendix

\section{Equations for the fluctuations}\label{Ap:pert}
Let us give for completeness some of the intermediate equations that were
skipped in the main body of the text for the sake of clarity.

\subsection{Sound Channel}
Here we find a set of 5 second order equations for the fluctuations that enter the gauge invariant expressions given in  \eqref{invsound}
\beqa
H_{tt}''&+&\ln'\!\left(\!\frac{c_T^2c_X^p}{c_R}\!\right)\!H_{tt}'-\ln'(c_T)H_{ii}'-c_R^2\left(\frac{\o^2}{c_T^2}H_{ii}+\frac{q^2}{c_X^2}H_{tt}+2\frac{q\o}{c_T^2}H_{tz}\right)\label{apHt}\\
&-&\frac{2}{p}c_R^2\frac{\d\pp}{\d\phi}\varphi=0 \label{pert-tt}\\
H_{tz}''&+&\ln'\!\left(\!\frac{c_X^{p+2}}{c_Tc_R}\!\right)\!H_{tz}'+\frac{c_R^2}{c_X^2}q\o H_{aa}=0 \label{pert-tz}\\
H_{zz}''&+&\ln'\!\left(\!\frac{c_Tc_X{^{p+1}}}{c_R}\!\right)\!H_{zz}'+\ln'(c_X)(H_{aa}'-H_{tt}')+c_R^2\!\left(\frac{\o^2}{c_T^2}H_{zz}+2\frac{q\o}{c_T^2}H_{tz}+\frac{q^2}{c_X^2}(H_{tt}-H_{aa})\!\right)\nonumber\\
&+&\frac{2}{p}c_R^2\frac{\d\pp}{\d\phi}\varphi=0 \label{pert-zz}\\
H_{aa}''&+&\ln'\!\left(\!\frac{c_Tc_X{^{2p-1}}}{c_R}\!\right)\!H_{aa}'+\ln'(c_X{^{p-1}})(H_{zz}'-H_{tt}')+c_R^2\!\left(\!\frac{\o^2}{c_T^2}-\frac{q^2}{c_X^2}\!\right)\!H_{aa}\nonumber\\
&+&\frac{2(p-1)}{p}c_R^2\frac{\d\pp}{\d\phi}\varphi= \label{pert-aa}0.\label{apHa}
\\
\varphi''&+&\ln'\!\left(\!\frac{c_Tc_X^p}{c_R}\!\right)\!\varphi'+c_R^2\left(\frac{\o}{c_T^2}-\frac{q^2}{c_X^2}\right)\varphi+\med \phi'(H_{ii}-H_{tt}))'-\frac{1}{\b}\frac{\d^2\pp}{\d\phi^2}\varphi =0 \label{pert-dil}
\eeqa
Additionaly there are three constraints associated with the gauge fixing condition $h_{\mu r}=0$
\beqa
&&H_{ii}'+\ln'\!\left(\!\frac{c_X}{c_T}\!\right)\!H_{ii}+\frac{q}{\o}H_{tz}'+2\frac{q}{\o}\ln'\!\left(\!\frac{c_X}{c_T}\!\right)\!H_{tz}+\b\phi'\varphi=0 \label{pert-tr}\\
&&H_{tt}'-\ln'\!\left(\!\frac{c_X}{c_T}\!\right)\!H_{tt}+\frac{\o}{q} \frac{c_X^2}{c_T^2}H_{tz}'-H_{aa}'-\b\phi'\varphi=0 \label{pert-zr}\\
&&\ln'(c_Tc_X{^{p-1}})H_{ii}'-\ln'(c_X{^p})H_{tt}'+c_R^2\!\left(\frac{\o^2}{c_T^2}H_{ii}+2\frac{q\o}{c_T^2}H_{tz}+\frac{q^2}{c_X^2}(H_{tt}-H_{aa})\!\right)\nonumber\\
&&\hspace{1cm}-\b\phi'\varphi'+c_R^2\frac{\d\pp}{\d\phi}\varphi= \label{pert-rr}0
\eeqa
It is straightforward to  check that together with equations \eqref{pert-tt}-\eqref{pert-dil} 
this system of 8 equations is not overdetermined, and one can construct easily
3  linear combinations that vanish identically ``on shell" (that is, modulo the equations of motion).

\subsection{Shear Channel}\label{Ap:shear}

Here we obtain two second order equations and one constraint
\beqa
H_{ta}''&+&\ln'\!\left(\!\frac{c_X^{p+2}}{c_Tc_R}\!\right)\!H_{ta}'-q  \frac{c_R^2}{c_X^2}\left(qH_{tz}+\o H_{za}\right)=0,\label{tshear}\\
H_{za}''&+&\ln'\!\left(\!\frac{c_Tc_X^p}{c_R}\!\right)\!H_{za}'+\o  \frac{c_R^2}{c_T^2}\left(qH_{tz}+\o H_{za}\right)=0,\label{zshear}
\\
qH_{za}'&+&\o \frac{c_X^2}{c_T^2}H_{ta}'=0.\label{shearconst}
\eeqa
which are again differentially linearly dependent.

\end{document}